\documentstyle [12pt] {article}
\begin{document}

\title{ The 2-Parametric Extension of $h$ Deformation\\
of $GL(2)$, and The Differential Calculus\\
on Its Quantum Plane \\ }
\author{ Amir Aghamohammadi }
\date { }
\maketitle
\begin{center}
{\it{Institute for studies in Theoretical Physics and Mathematics
\\ P.O.Box:19395-1795  Tehran Iran}}\\
\end{center}
\begin{center}
{\it{ and}} \\
\end{center}
\begin{center}
{\it Alzahra University, Physics Department,\\ P.O.Box:19395-3199
Tehran Iran}\\
\end{center}
\vspace {10mm}
\begin{abstract}
{We present an alternative 2-parametric deformation
 $ GL(2)_{h,h'} $ ,
and construct the differential calculus on the quantum plane on which
this quantum group acts. Also we give a new deformation of the two
dimensional Heisenberg algebra.}\\ \\
\end{abstract}
\newpage
\noindent
{\large \bf I. Introduction \\ }
Recently quantum matrices in two dimensions, admitting left and right
quantum spaces, are classified$^1$. They fall into two families. One
of them is the 2-parametric extension of $q$ deformation of $GL(2)$,
which is well studied$^2$. There is an alternative case which
its $R$ matrix is given in reference 1, and we denote it by
$R_{h,h'}$. In this paper we
construct the quantum group associated with the $R_{h,h'}$.
On the other hand, it is also shown that$^3$,
up to isomorphism, there exist just two quantum deformation of
$GL(2)$ which admit a central determinant, the well known $q$
deformation
and recently constructed $h$ deformation.

The 2-parametric defomation  $GL(2)_{q,p} $ is studied in ref. 2. $R$
matrix associated with this quantum group which solves quantum Yang
Baxter equation

\begin{equation}
R_{12}R_{13}R_{23}=R_{23}R_{13}R_{12}
\end{equation}
is
\begin{equation}
R_{qp}=\left ( \begin{array}{llll} q&\ 0&\ 0&0 \\0&\ 1&\ 0&0 \\
0&q-p&q/p&0 \\ 0&\ 0&\ 0&q \end{array}\right )
\end{equation}

Using the above $R$ matrix and the method developed in ref.4,
the algebra
of the elements of quantum matrix
$T=\left ( \begin{array}{ll} a & b \\
 c & d  \end{array} \right )$  can be obtained.
For the two parametric case, the quantum determinant is

\begin{equation}
{\cal D}=ad-pbc=ad -qcb=da-p^{-1}cb=da-q^{-1}bc
\end{equation}

The crucial difference with the one parametric case, is that
the quantum determinant is not central but satisfies the following
relations

\begin{equation}
[{\cal D},a]=[{\cal D},d]=0,\ \ \ \ q{\cal D}=pb{\cal D}
,\ \ p{\cal D}c=qc{\cal D}
\end{equation}
there is an alternative  $R$ matrix
\begin{equation}
R=\left ( \begin{array}{llll}
1 & -h' & h'&hh' \\
0 & \ 1 & 0 &-h \\
0 & \ 0 & 1 &\ h \\
0 &\  0 & 0 &\ 1
\end{array} \right )
\end{equation}
which solves (1). The algebra of polynomials on the quantum $GL(2)$
associated with the above $R$ matrix
for the special cases $h=h'=1$  and $h=h'$ are studied in
references 5 and 6. The universal enveloping algebra $U_h(sl(2))$
has also  been constructed$^7$ and the quantum de Rham complexes
associated with $h$
deformation of $sl(2)$ is given in ref. 8.

The quantum groups which are associated with two matrices
$R$ and $R'=(S\otimes S)R(S\otimes S)^{-1}$ are equivalent. This is
the case for the matrix $R$ as given in eq. (5) when $h$ equals $h'$,

\begin{equation}
R_{h=h'=1}=(S\otimes S)R_{h=h'}(S\otimes S)^{-1},\ \ \ S=\left (
\begin{array}{ll} h^{-1/2}&0 \\ 0&h^{1/2} \end{array} \right )
\end{equation}

So for the special case $h=h'$, all the Hopf algebras for $h \ne 0$
are isomorphic to the case $h=1$. Thus $h$ is not a continous
parameter of deformation.
However for the general case $h\ne h'$, there is no such $S$ which
simultanousely fixes $h$ and $h'$ to one. Of course one can always
fix $h$ to one, but since we are interested in the classical limit
$h\to 0 ,\ \ h'\to 0 $ , we do not fix it to be one.
In this paper we study 2-parametric deformation of $GL(2)$,
$GL(2)_{h,h'}$, the quantum plane on which it acts and differential
calculus on that plane.

\newpage

\noindent
{\large \bf II. The Algebra of Functions \\ }

Following the method of ref. 4 and using the $R$ matrix (5),
we arrive at the commutation relations of the quantum matrix
$ T= \left ( \begin{array}{ll}a&b \\ c&d \end{array} \right ) $
\begin{equation}
[a,c]=hc^2 \ \ \ \ [b,c]=hcd+h'ac \ \ \ \ [a,b]=h'({\cal D} -a^2)
\end{equation}
\begin{equation}
[d,c]=h'c^2 \ \ \ \ [a,d]=hcd-h'ca \ \ \ \ [d,b]=h({\cal D} -a^2)
\end{equation}
where ${\cal D}$ is
\begin{equation}
{\cal D}=ad-cb-hcd=ad-bc+h'ac
\end{equation}
This algebra can then be made a bialgebra $A_{h,h'}(2)$ by definition
of co-product and co-unit
\begin{equation}
\Delta (T_{ij})=T_{ik}\otimes T_{kj},\ \ \
\epsilon (T_{ij})=\delta_{ij} \end{equation}
So
\begin{equation}
\Delta \left ( \begin{array}{ll} a&b \\ c&d \end{array} \right )
 =\left ( \begin{array}{ll}{a\otimes a+b\otimes c}&{a\otimes b+
b\otimes d}\\ {c\otimes a+ d\otimes c}& {c\otimes b +d\otimes d}
 \end{array} \right )
\end{equation}
\begin{equation}
\epsilon \left ( \begin{array}{ll} a&b \\ c&d \end{array} \right )
=\left ( \begin{array}{ll} 1&0 \\ 0&1 \end{array} \right )
\end{equation}
For turning $A_{h,h'}(2)$ into a Hopf algebra we shoud be able to
write down explicitly the inverse of the quantum matrix $T$.
${\cal D}$, which is defined in (9) is quantum determinant and by
direct verification it can be shown that
\begin{equation}
{\cal D}(TT')={\cal D}(T){\cal D}(T'),\ \ \ if \ \ [T_{ij},T'_{kl}]=0
\end{equation}
\begin{equation}
\Delta ({\cal D})={\cal D }\otimes {\cal D}
\end{equation}
\begin{equation}
\epsilon ({\cal D})=1
\end{equation}
For the general case ${\cal D}$ is not central and its commutation
relations with elements of $T$ is

\begin{equation}
[{\cal D},a]=[d,{\cal D}]=(h'-h){\cal D}c
\end{equation}

\begin{equation}
[{\cal D},c]=0,\ \ \ \ [{\cal D},b]=(h'-h)({\cal D}d-a{\cal D})
\end{equation}

This is reminiscent of the other 2-parametric deformation,
$GL(2)_{qp}$, where the quantum determinant is not central. If
${\cal D}\ne 0 $ one extends the algebra by an inverse of ${\cal D}$
which obeys

\begin{equation}
{\cal D}{\cal D}^{-1}={\cal D}^{-1}{\cal D}=1
\end{equation}
from which it follows that:
\begin{equation}
[{\cal D}^{-1},a]=[d,{\cal D}^{-1}]=(h-h'){\cal D}^{-1}c
\end{equation}

\begin{equation}
[{\cal D}^{-1},c]=0,\ \ \ \
[{\cal D}^{-1},b]=(h-h')(d{\cal D}^{-1}-{\cal D}^{-1}a)
\end{equation}

It is easy to show that

\begin{equation}
MT=TM'={\cal D}\left (\begin{array}{ll} 1&0\\ 0&1 \end{array}\right )
\end{equation}

where

\begin{equation}
M\ = \left ( \begin{array}{ll} d+hc& -b+h(d-a)+h^{2}c
\\ -c& \ \ \ \ a-hc \end{array} \right )
\end{equation}

\begin{equation}
M'= \left ( \begin{array}{ll} d+h'c & -b+h'(d-a)+h'^{2}c
\\ -c& \ \ \ \ a-h'c \end{array} \right )
\end{equation}

So a consistent definition of inverse can be given by

\begin{equation}
T^{-1}={\cal D}^{-1}M=M'{\cal D}^{-1}
\end{equation}

Clearly for the case $h=h'$, $M$ and $M'$ coincide and determinant
${\cal D}$ is in the center of algebra.
The quantum group
$GL(2)_{h,h'}$ is defined as the Hoph algebra obtained from the
bialgebra $A_{h,h'}(2)$ extended by the element ${\cal D}^{-1}$
and the antipode given by
\begin{equation}
S(T)=T^{-1}= \left ( \begin{array}{ll} d+h'c& -b+h'(d-a)+h'^{2}c
\\ -c& \ \ \ \ a-h'c \end{array} \right ){\cal D}^{-1}
\end{equation}
\\
{\large \bf III. Differential Calculus on The Quantum Plane \\ }

In this section we  will construct a covariant differential
calculus on the
quantum plane.
General formalism for constructing
differential calculus on the quantum plane has been given by Wess
and Zumino$^9$. In this paper we use the formalism of ref. 10.
Consider quadratic relations between coordinates $x_i$ of a
non-commutative space
\begin{equation}
C^{ij}_{kl}x^kx^l=0
\end{equation}
Introducing partial derivatives
\begin{equation}
\partial _i=\partial /\partial x_i ,\ \ \ \ (\partial _ix^k)=
\delta _i^k
\end{equation}
and assuming deformed Leibniz rule for partial derivatives
\begin{equation}
\partial _i(fg)=(\partial f)g+O^j_i(f)\partial _j g, \ \ \ where \ \
\ \ O^j_i(x^k)=Q^{kj}_{in}x^n
\end{equation}
one arrives at
\begin{equation}
\partial _ix^k=\delta _i^k+Q^{km}_{in}x^n\partial _m.
\end{equation}
If we now differentiate the commutation relations (26)
\begin{equation}
0=\partial _kC^{rs}_{ij}x^ix^j= C^{rs}_{ij}(\delta ^i_k \delta ^j_n
+Q^{ij}_{kn})x^n,
\end{equation}
Since there are no linear relation among the variables $x^i$, we have
the following commutation relations
\begin{equation}
C^{rs}_{ij}(\delta ^i_k \delta ^j_n
+Q^{ij}_{kn})=0.
\end{equation}
By defining an exterior derivative
\begin{equation}
d=\xi ^i \partial _i
\end{equation}
which satisfy the undeformed Leibniz rule and the co-boundary
condition
\begin{equation}
d(fg)=(df)g+fdg
\end{equation}      \begin{equation}
d^2f=0
\end{equation}
one can obtain the following commutation relations
(see ref. 10 for more details)
\begin{equation}
x^ix^j=(\delta ^i_k \delta ^j_l-C^{ij}_{kl})x^kx^l=B^{ij}_{kl}x^kx^l
\end{equation}
\begin{equation}
x^i\xi ^j=Q^{ij}_{kl}\xi ^kx^l
\end{equation}
\begin{equation}
\xi^i\xi ^j=-Q^{ij}_{kl}\xi ^k\xi ^l
\end{equation}
\begin{equation}
\partial _ix^j=\delta ^j_i+Q^{jn}_{im}x^m\partial _n
\end{equation}                       \begin{equation}
\partial _i\xi^j=(Q^{-1})^{jn}_{im}\xi ^m\partial _n
\end{equation}                       \begin{equation}
\partial _i\partial _j=S^{mn}_{ij}\partial _m\partial _n
\end{equation}
with the following consistency relations
\begin{equation}
(\delta ^r_i \delta^s_j+B^{rs}_{ij})(\delta ^i_k \delta ^j_n
+Q^{ij}_{kn})=0
\end{equation}
\begin{equation}
(\delta ^k_m \delta^l_n+Q^{kl}_{mn})(\delta ^m_i \delta ^n_j
-S^{nm}_{ji})=0
\end{equation}
and $Q$ should satisfy the Braid group equation
\begin{equation}
Q_{12}Q_{23}Q_{12}=Q_{23}Q_{12}Q_{23}
\end{equation}
if we choose
\begin{equation}
Q=\left ( \begin{array}{llll}
1 & -h' & h'&hh' \\
0 & \ 0 & 1 &\ h \\
0 & \ 1 & 0 &- h \\
0 &\  0 & 0 &\ 1
\end{array} \right ),\ \ \ B=Q,\ \ S=PQP
\end{equation}
where $P$ is the permutation matrix, then all the
consistency relations (41-43) are satisfied.
By inserting (44)in (35) we obtain the relation of quantum plane
$R_h(2)$ with the coordinates $x$ and $y$
\begin{equation}  [x,y]=hy^2             \end{equation}
Similarly one can obtain the relations of dual quantum plane
$R^*_{h'}(2) $ with the coordinates $\xi$ and $\eta $
\begin{equation} \eta ^2=\xi \eta +\eta \xi=0,\ \ \ \
\xi^2=h'\xi \eta \end{equation}
This means that $T$ acts on the $h$ plane and $h'$ exterior plane.
The relations between coordinates of quantum plane and its dual are
\begin{equation}
[x,\eta ]=h\eta y,\ \ \ [x,\xi ]=h'(x\eta -\xi y)
\end{equation}
\begin{equation}
[y,\eta ]=0,\ \ \ \ \ [y,\xi ]=-h\eta y
\end{equation}
and the deformed Leibniz rule is given by:
\begin{equation}
[\partial _x,x]=1-h'y\partial _x,\ \ \ \
[\partial _y,y]=1-hy\partial _x
\end{equation}
\begin{equation}
[\partial _x,y]=0\ \ \ \ [\partial _y,x]=h'x\partial _x
+hh'y\partial _x+hy\partial _y
\end{equation}
and for the derivatives we have
\begin{equation}
[\partial _x,\partial _y]=h'\partial _x^2
\end{equation}
Finally to complete the set of relations we give the relations
among $\partial _i$ and $\xi^k$
\begin{equation}
[\partial _x,\xi ]=-h'\eta \partial _x,\ \ \ \
[\partial _y,\eta ]=-h\eta \partial _x
\end{equation}
\begin{equation}
[\partial _x,\eta ]=0\ \ \ \ [\partial _y,\xi]=h'\xi \partial _x
+hh'\eta \partial _x+h\eta \partial _y
\end{equation}
\\
\noindent
{\large \bf IV. Deformed Heisenberg Algebra \\ }

Now we will give a new deformed two dimensional Heisenberg algebra.
It is interesting to note that for the general case (and also for
the one parametric case $h=h'$)
identifying $\partial _x$ and $\partial _y$
with $ip_x$ and $ip_y$ is not compatible with the hermiticity
of coordinates and momenta ( see (51) ).
To identify $\partial _x$ and $\partial _y$
with the momenta $ip_x$ and $ip_y$, one must care about
hermiticity of coordinates and momenta. This can be done by taking
$h$ as a pure imaginary parameter and $h'=-h$. Then
the hermiticity of $x,y,p_x$ and $p_y$ are compatible with the
relations (50-52).
The final form of the deformed
Heisenberg algebra is
\begin{equation}
[p_x,x]=-i+hyp_x \ \ \ \ \ \ [p_y,y]=-i-hyp_x
\end{equation}
\begin{equation}
[p_x,x]=0\ \ \ \ \ \ \ \ [p_y,x]=-hyp_x-h^2yp_x+hyp_y
\end{equation}
\begin{equation}
[p_x,p_y]=-hp_x^2\ \ \ \ \  [x,y]=hy^2
\end{equation}
This gives a deformed Heisenberg algebra which can be used to study a
two dimensional quantum space.

One of the intersting problems is constructing $U(gl(2))_{h,h'}$.
In the case of $q$ deformation, universal enveloping algebra of
multiparametric case  can be obtained,
by simply twisting$^{11}$, but for the $h$ deformation,
it should be clarified how to  multiparametrize
the universal enveloping algebra.\\

\noindent
{\bf Acknowledgement} \\

The author wishes to thank F. Ardalan, S. Rouhani and
 A. Shafiei Dehabad for valuable
discussions.\\ \\

\noindent
{\large References \\ }

\begin{enumerate}
\item H. Ewen, O. Ogievetsky, J. Wess, Lett. Math. Phys. {\bf 22},297
(1991).
\item A. Schirrmacher, J.Wess, B. Zumino, Z. Phys. C {\bf 49}, 317
(1991); O. Ogivetsky, J. Wess, Z. Phys. C {\bf 50},123
(1991); V. K. Dobrev, J. Math. Phys. {\bf 33} 3419 (1992).
\item B A Kupershmidt, J. Phys. A {\bf 25} L1239 (1992).
\item L. D. Faddeev, N. Yu. Reshetikhin, L. A. Takhtajan, Leningrad
Math. J. {\bf 1}, 193 (1990).
\item E. Demidov, Yu. I. Manin, E. E. Mukhin, D. V. Zhdanovich,
preprint RIMS-701 (1990).
\item S. Zakrzewski, Lett. Math. Phys. {\bf 22}, 287 (1991).
\item Ch. Ohn, Lett. Math. Phys. {\bf 25}, 89 (1992).
\item V. Karimipour, Sharif Univ. Preprint (1993).
\item J. Wess, B. Zumino, Nuclear Phys. B {\bf 18}, 302 (1990).
\item J. Schwenk, Proceeding of The Argonne Workshop (1990).
\item N. Yu. Reshetikhin, Lett. Math. Phys. {\bf 20}, 331 (1990).

\end{enumerate}
\end{document}